\documentstyle[12pt,aasms4]{article}
\begin{document}

\title{ ANOMALOUS COOLING OF THE MASSIVE WHITE DWARF IN U GEMINORUM
FOLLOWING A NARROW DWARF NOVA OUTBURST $^*$}

\author{Edward M. Sion$^{1}$, F.H. Cheng$^{1,2}$, Paula Szkody$^{3}$,
Warren Sparks$^{4}$, Boris G\"{a}nsicke$^{5}$,\\ 
Min Huang$^{6}$, \& Janet Mattei$^{7}$ }
\vspace*{0.5cm}

\noindent $^{1}$Department of Astronomy \& Astrophysics,  Villanova 
University, Villanova, PA 19085.

E-mail: emsion@ucis.vill.edu, fhcheng@ucis.vill.edu.\\
$^{2}$ Center for Astrophysics, University of Science and Technology of
China, Hefei,

Anhui 230026, People's Republic of China.\\
$^{3}$ Department of Astronomy,  University of Washington,
Seattle, WA 98195.

E-mail: szkody@astro.washington.edu\\
$^{4}$ X-3, Los Alamos National Laboratory, Los Alamos, NM 85723.
		wsparks@lanl.gov\\
$^{5}$ Universitat-Sternwarte G\"{o}ttingen, Geismarlandstra$\beta$e 11, 
D-37083, Germany. 

boris@uni-ew.gwdg.de\\
$^{6}$ Arthur Andersen Inc., New York, NY 10103,
min.a.huang@arthurandersen.com\\
$^{7}$ American Association of Variable Star Observers, Cambridge, MA.
		aavso@aavso.org\\[2cm]
\noindent$^{*}$ Based on observations with the NASA/ESA Hubble Space 
Telescope,
obtained at the Space Telescope Science Institute, which is operated by
the Association of Universities for Research in Astronomy, Inc., under
NASA contract NAS5-26555.\\
\newpage
\begin{center}
{\bf Abstract}\\[2cm]
\end{center}

We obtained Hubble GHRS medium resolution (G160M grating) phase-resolved
spectroscopic observations of the prototype dwarf nova U Geminorum
during dwarf nova quiescence, 13 days and 61 days following the end of a
narrow outburst.  The spectral wavelength ranges were centered upon
three different line regions: N V (1238\AA, 1242\AA), Si~III (1300\AA)
and He II (1640\AA).  All of the quiescent spectra at both epochs are
dominated by absorption lines and show no emission features.  The Si III
and He II absorption line velocities versus orbital phase trace the
orbital motion of the white dwarf but the N~V absorption velocities
appear to deviate from the white dwarf motion.  We confirm our
previously reported low white dwarf rotational velocity, V sin$i$= 100
km s$^{-1}$.  We obtain a white dwarf orbital velocity semi-amplitude 
K$_{1}$=107 km s$^{-1}$.  Using the $\gamma$ velocity of Wade (1981) we
obtain an Einstein redshift of 80.4 km s$^{-1}$ and hence a carbon core
white dwarf mass of $\sim$1.1  M$_\odot$.  We report the first subsolar
chemical abundances of C and Si for U Gem with C down by 0.05 with
respect to the Sun, almost certainly a result of C depletion due to
thermonuclear processing.  This C-depletion is discussed within the
framework of a weak TNR, contamination of the secondary during the
common envelope phase, and mixing of C-depleted white dwarf gas with
C-depleted matter deposited during a dwarf nova event.  Remarkably the
T$_{eff}$ of the white dwarf 13 days after outburst is only 32,000K,
anomalously cooler than previous early post-outburst measurements.
Extensive cooling during an extraordinarily long (210 days) quiescence
followed by accretion onto an out-of-equilibrium cooled degenerate could
explain the lower T$_{eff}$.

\section{Introduction}

The dwarf nova U Geminorum undergoes both wide ($\sim$ 14 days) and
narrow ($\sim$ 4-7 days) outbursts during which it is expected that
differing amounts of mass and angular momentum accretion occur onto the
white dwarf.  Therefore in a continuing effort to elucidate the
tangential accretion physics, in particular the actual mechanism of
accretional heating, e.g. shear mixing, compression and irradiation, it
is of interest to compare the response of the white dwarf to outbursts
of different lengths, for example, a wide outburst versus a narrow
outburst.  The differential heating affect of a normal outburst versus a
superoutburst has already been demonstrated for the case of the white
dwarf in VW Hydri (G\"{a}nsicke \& Beuermann 1996; Sion et al. 1996). 
We obtained high resolution GHRS
observations of U Gem during the quiescence following a narrow outburst
which allow us to do just that.  Moreover we obtained these spectra at
the orbital quadratures to maximize the velocity displacements of the
lines, delineate different regions of line formation and estimate the
mass of the white dwarf from its gravitational redshift.  In this paper
we report the results of our experiment and interpret the results
comparatively with the results and predictions of earlier investigations
of U Gem.\\

\section{ HST GHRS Far Ultraviolet Observations}

Upon notification of the onset of an outburst of U Gem by AAVSO
observers, we obtained two sets of GHRS observations of U Gem during the
following quiescence, the first set on 1995 October 10 (Obs1) and the
second set on 1995 November 27 (Obs2).  The temporal placement of the
observations with respect to the narrow outburst is shown in figure 1
where we present the AAVSO light curve data for the outburst and the
following quiescence.  The first set of observations took place 13 days
after outburst while the second dataset was obtained 61 days after
outburst.  The observations for both observations were carried out in
the ACCUM mode with the D2 detector of GHRS and the G160M disperser.
Three wavelength regions were covered by the observations: the N V
region (1219\AA\ to 1255\AA), the Si III region (1269\AA\ to 1304\AA)
and the He II region (1616\AA\ to 1648\AA), with a resolution of 0.25
\AA.  The wavelength scale has an accuracy of $\sim$0.10 \AA\ Since the
objective of our line formation study was to delineate the white dwarf
photosphere in quiescence and obtain maximum velocity displacement and
mass information for the white dwarf, the observations were obtained
close to the quadrature points of the orbit. \\

\begin{center}
\begin{tabular}{lccrcrc}
\multicolumn{7}{c}{Table 1. HST GHRS Observations of U Gem}\\ 
\hline\hline
\multicolumn{7}{c}{Quiescence Obs. No. 1:  95/10/10}\\
\hline
Ion&Start Time&T$_{exp}$ (s)&Start (MJD)&$\Phi$&End (MJD)&$\Phi$\\
\hline
N V     & 18:36:31&  1767&  50000.77536378 &  0.40&    50000.80212738& 
0.55\\
Si III  & 20:09:49& 1767&  50000.84015545 &  0.76&    50000.86692050 & 
0.92\\
He II   & 21:46:05& 1767&  50000.90700730 &   0.14&    50000.93377235&  
0.29\\
He II   & 23:22:35&  408&  50000.97402119&    0.52&   50000.98018583&  
0.56\\
\hline
\multicolumn{7}{c}{Quiescence Obs. No. 2:  95/11/27}\\ [0.02in]
\hline
N V   &   15:02:37&  1767&  50048.62682216   & 0.897&50048.65358866&0.04 \\
Si III &  16:38:29&  1767&  50048.69339624  &  0.27 &50048.72015984&0.42 \\
He II  &  18:14:59&  1767&  50048.76041157 &   0.65 &50048.78717517&0.80\\
He II  &  19:51:29&   408&  50048.82742402& 0.02 &50048.83358866&0.06\\
\hline
\end{tabular}
\end{center}

A detailed observing log of the observations is given in Table 1 where
we tabulate for each ion wavelength region the start time of the
observation, the total exposure time in seconds, the start and end times
in modified Julian Date (MJD) and the orbital phase at the start and end
times of each exposure.  For the phasing we adopted the orbital
ephemeris of Marsh et al.  (1990) where phase 0.0 corresponds to
inferior conjunction of the secondary star; viz.,   \\

[HJD=32437638.82325 + 0.1769061911]\\

In this phase convention, the white dwarf would have maximum positive
velocity at phase 0.75 and maximum negative velocity at phase 0.25.
While the Si III and He II velocities during obs1 and obs2 are
consistent with the expected motion of the white dwarf, the N V
velocities are inconsistent with this motion.  Therefore, the N V
absorption features cannot be associated with the Einstein-redshifted
rest frame of the white dwarf photosphere.

In Table 2 we present measurements of the strongest absorption features
in the three wavelength regions.  For the N V doublet and the five
individual members of the Si III multiplet, we have tabulated the
average of the individual velocities.

\begin{center}
\begin{tabular}{lccrcrcc}
\multicolumn{8}{c}{Table 2. Line measurements}\\ \hline\hline
Ion &$\lambda$ (rest)&Obs1&$<\Phi>$&$<$V1$>$&Obs2&$<\Phi>$&$<$V2$>$\\
\hline
NV   &   1238.821 &      1238.98&  0.48 &   +50     &1239.62&  0.96&   
+202\\
&  1242.804 &      1243.05&        &          &1243.67&             &\\
Si III&  1294.545 &      1295.56&  0.84  & +251     &1294.87 & 0.35& +75 \\
						&  1296.726 &      1297.93   &     &          &1297.08 &     &     
				\\
						&  1298.946 &      1299.94   &     &          &1299.23&&\\
						&  1301.149 &      1302.23&        &          &1301.49 &&          
					\\
						&  1303.322 &      1304.46    &    &&          1303.65&&\\
He II &  1640.414 &      1640.78    &0.22&  +56     &1641.52&0.73&
+191\\
\hline
\end{tabular}
\end{center}

\section{Surface Temperatures and Chemical Abundances During Quiescence}

It is clear that with only three GHRS settings covering different 35 
\AA\
wavelength regions and different orbital phase ranges,
synthetic spectral fitting will be less accurate
than if the entire far UV spectrum were available.  This disadvantage of
the limited continuum is offset slightly by the detailed line profile
information we have available. We have assumed no temporal changes
occurred
within each set of HST observations and fitted synthetic spectra
to the three spectral regions simultaneously, for obs1 and obs2.

Our fitting attempt utilized both single temperature white dwarf models
as well as combined white dwarf plus accretion belt synthetic fluxes.
The details of our fitting procedure is the same as in our previous
analyses and for the sake of brevity will not be repeated here (see Sion
et al. 1996; Cheng et al. 1997).

Our best fitting single temperature white dwarf model yielded
the values for obs1 and obs2 shown in Table 3.\\
\begin{center}
\begin{tabular}{lcc}
\multicolumn{3}{c}{Table 3. Results of White Dwarf Spectral Fitting}\\
\hline\hline
Parameter &    Obs1 (13 Days POB)  &    Obs2 (61 days POB)\\
\hline
log g     &        8.0               &         8.0   \\
T$_{wd}$ (10$^3$ K) &      32.2              &       30.0  \\
V$_{rot}$ (km s$^{-1}$)     &       100              &        120  \\
Abundances (in units of Solar)&&\\
C         &       0.05             &        0.05 \\
Si        &      0.4               &       0.4          \\
He        &      1.0               &       1.0       \\
Others    &       1.0              &         1.0             \\
\hline
\end{tabular}
\end{center}

The derived abundances are the first to indicate sub-solar values in the
accreted atmosphere of the U Gem white dwarf. The C and Si abundances
are significantly lower than the essentially solar abundances derived in
earlier HST and HUT analyses at lower spectral resolution (Cheng et al.
1997; Long et al. 1993; Long et al. 1996), a point we return to in
the concluding section. Note also that the magnitude of the heating and
cooling is also reduced compared with earlier analyses (see section 4).

Two temperature fits were also attempted subject to the earlier caveats
regarding the limited spectral coverage.  The best results were achieved
with white dwarf models having essentially the same temperatures as in
Table 1, in combination with a rapidly spinning belt with V$_{belt}$ 
= 3,300 km s$^{-1}$ and T$_{belt}$ = 45-50,000 K. While these fits 
yielded
slightly lower $\chi^2$ values than the single white dwarf models, we
found less agreement with the depths of the absorption line features,
especially the Si III photospheric lines which are fit nearly perfectly
by a slowly rotating single temperature white dwarf model.

\section{ The Gravitational Redshift Mass of the U Geminorum White Dwarf
and Its Implications}

	In Table 2 we list the rest wavelengths, the observed wavelength
measurements, the corresponding orbital phase at mid-exposure for each
ion, and the corresponding velocities of N V , Si III, and He II for
obs1 and obs2.  The shift $\Delta\lambda$ is defined as
$\lambda_{obs}$-$\lambda_{model}$.

The observed shifts for Si III and He II in both obs1 and obs2 are
consistent in phase with the expected motion of the white dwarf.
However the N V absorption shows a peculiar shift with V = 202 km
s$^{-1}$ near zero phase!  We regard the Si III as completely
photospheric in origin.  However, given that He II may also be
contributing from the same high temperature region as N V, (i.e.
T$_{eff} >$ 80,000K) it is unlikely to be entirely photospheric and is
therefore disregarded in the gravitational redshift determination.
Given that Si III is Einstein-redshifted in the rest frame of the white
dwarf and that its multiplet consists of 6 individually-resolved lines,
we took an average of the individual transitions as the true global
photospheric feature, and adopt it for a gravitational redshift
determination.  At mid-exposure phase 0.84 the Si III velocity is 251 km
s$^{-1}$ while at mid-exposure phase 0.35 the Si III velocity is 75 km
s$^{-1}$. 

Using Si III and assuming a sinusoidal relation to solve for K1, we find
it is 107  km s$^{-1}$.  At present there are two well-measured but
discordant values of the gamma velocity of U Gem, 84  km s$^{-1}$ (Wade
1981) and 43 km s$^{-1}$ (Friend et al. 1993).  Adopting the systemic
velocity of Friend et al.  (1993), viz., 43 km s$^{-1}$, we find that
the gravitational redshift of the white dwarf is 118 km s$^{-1}$.
Adopting the systemic velocity of Wade (1981), viz., 84 km s$^{-1}$, we
find that the gravitational redshift of the white dwarf is 77 km
s$^{-1}$.  We do not know which is closer to being correct so we will
take a mean of the two.  If we take this mean gamma velocity, we find
the resulting redshift is 99 km s$^{-1}$.   It is quite remarkable that
all three of these redshift values, when compared with the mass-radius
relation from the extensive grid of evolutionary models
by Wood (1996) for a carbon core, indicate a very massive white dwarf.
While we cannot rule out an O-Ne-Mg core for the U Gem degenerate, the
redshift yielded by using the Friend et al. value can almost certainly
be ruled out because it implies a mass exceeding the Chandrasekhar mass.
Even the mean value (62 km s$^{-1}$) yields a mass $\geq$ 1.2 M$_\odot$!
The Wade (1981) value however yields an entirely reasonable mass of 1.1
M$_\odot$, a result in agreement with the optical spectroscopic radial
velocity study of Stover (1981; see also Zhang and Robinson 1987) which
yielded a white dwarf mass of 1.18 M$_odot$.  Furthermore Webbink's
critical systematic re-determination of CV white dwarf masses yielded a
value for the U Gem white dwarf of $\sim$1.1 M$_\odot$.

\section{Heating and Cooling of the White Dwarf}

It is surprising that the white dwarf T$_{eff}$ values 13 days and 61
days after outburst are cooler T$_{eff}$ measurements at comparable
times after outburst than previous studies.  This is supported by a
lower flux level of our GHRS spectra (by 4$\times10^{-14}$) compared
with all other post-outburst temperature measurements at comparable
times in quiescence (e.g.  Sion et al. 1994; Long et al. 1993, 1994,
1995).  Our observations and the FOS observations of Long et al. 1994
were both obtained following a narrow outburst of U Geminorum.  Since we
expect that the amount of heating of the white dwarf and the subsequent
rate of cooling should be similar following the same types of outburst,
then it is clear that the white dwarf T$_{eff}$ 13 days POB (32,000K) is
considerably lower than the value (39,000K) measured 13 days after the
narrow outburst of U Gem reported by Long et al.  (1994).  We believe
the T$_{eff}$ difference is real and is almost certainly related to the
extraordinarily long quiescence experienced by U Gem in 1994/95 , which
was ended after 210 days by a wide outburst in April 1995 , then a short
68 day quiescence followed by the narrow outburst preceding our
observations (see Fig.1 ).  The normal quiescent interval of U Gem is
118 days (Szkody \& Mattei 1984).  Long et al.  (1996) reported a
T$_{eff}$ of 29,000K 185 days into the long 210 day quiescence.  Since
that quiescence lasted another 25 days, it is even possible that some
additional cooling of the white dwarf took place.  Therefore, if the
white dwarf had cooled down to 27-29,000K, the compressional heating
calculations of Sion (1995) at a rate $\dot{M}$ = 10$^{-8}$ M$_\odot$
yr$^{-1}$ for 7 days would predict a peak heating of only $\sim$35,000K
at the exact end of the outburst and a subsequent cooling down to
$\sim$27,000K which is not too far below the estimated T$_{eff}$ at 185
days POB by Long et al.  (1994).  In this scenario, the long quiescence
could have disrupted a normal time-averaged "equilibrium" between
accretional heating of the upper envelope and cooling by radiation.  The
normal (average) equilibrium would be re-established only after a
sufficient number of dwarf nova cycles.

On the other hand, the long (210 day) quiescence could have led to a
complete or nearly complete spin down of differentially rotating white
dwarf surface layers (e.g. an accretion belt; see Cheng et al. 1997).
Hence, subsequent accretion events during the following wide and narrow
outbursts would have deposited mass and energy with a percentage-wise
greater dissipation in the boundary layer, thus resulting in a greater
proportion of the accretion energy being released at soft X-ray/EUV
wavelengths.  This may account for the lower surface temperature we
observe post-outburst and would directly manifest a dependence of the
heating of the white dwarf on the accreting star's short term angular
momentum history.\\

\section{Implications of the White Dwarf Chemical Abundances}

If the white dwarf accretes solar or nearly solar C, then C must rapidly
gravitationally settle out of it's atmosphere.  This leads to two major
conflicts.  Why don't the other metals also gravitationally settle out
and why is the C abundance the same sub-solar value at 13 and 61 days
after the dwarf nova outburst?  For example, ongoing accretion of a
solar mix of gas during quiescence would quickly replenish
diffusion-depleted C. There is another possible solution: an ancient
thermonuclear runaway (TNR).

It is fully expected that U Gem and all other dwarf novae will undergo
(and have undergone in the past) a TNR when the WD has accumulated
sufficient hydrogen-rich material (Starrfield, Sparks and Shaviv 1988
).
During a TNR, C proton captures to form N. If the C abundance is larger
than solar, then a strong TNR results, leading to a nova outburst
(Starrfield 1995) .  This C overabundance comes from the accreted
material mixing with the white dwarf's core material (Starrfield,
Truran, Sparks and Kutter 1972) Thus, most observations of novae end up
with an overabundance of C, even though much of the C has been processed
to N. If there is no mixing with the core, then in most cases the TNR
will be weak.  The notable exception is a rapidly accreting, very high
mass WD ($\sim$1.35 M$_\odot$), which is associated with recurrent novae
(Sparks, Kutter, Starrfield and Truran, J. 1990) .  For a more slowly
accreting WD (as in the case of a dwarf novae, or a less massive WD),
the TNR will be weaker, little or no material will be ejected during the
outburst, and a large common envelope will form.  Although some of the
common envelope may be ejected, a fraction will be deposited on the
secondary and a similar fraction will be consumed by the white dwarf's
rekindled H shell source.  This shell source will leave a He-rich
material layer enriched in N and depleted in C. This He layer should
prevent core material from being mixed up, thus leading to subsequent
weak TNRs.  Later dwarf novae will deposit C depleted material due to
the TNR and common envelope to be mixed with even stronger C depleted WD
material due to the TNR and the remnant H shell burning source.  Thus C
will be depleted and the N will be enhanced in both the accreted
material and the WD's surface material.  We did not cover any
photospheric N lines in our GHRS setting with which to obtain an N
abundance but our prediction is that N should be overabundant in U Gem.

A number of our HST observations of U Gem support or are, at least,
consistent with this scenario.  First, the very slow WD's rotation
velocity and very low C abundance are indicators that not much material
has been accreted since the last TNR. Both the C abundance and the
rotational velocity will increase with the amount of accreted material.
Second, the large WD mass means that the amount of accreted mass needed
to trigger TNR will be small.  The small accreted mass implies that the
accretion timescale will also be short.  A short accretion timescale
works against two of the four proposed mixing mechanisms:
accretion-driven shear mixing and diffusion (Livio 1993).  The weaker
TNR from the low initial C abundance hinders the other two mechanisms
(convection-driven shear mixing and undershooting) from penetrating the
remnant He layer.  If this scenario is correct it means that U Gem and
probably other dwarf novae are massive WDs increasing in mass with the
possibility of becoming a SNI.

\section{Concluding Remarks}

Using GHRS G160M spectroscopic observations, we have uncovered surprises
in further characterizing the physical characteristics of the white
dwarf in U Gem and its response to heating by the dwarf nova outburst.
The markedly lower elevation of white dwarf surface temperature we have
measured, compared to previous observations, is probably related to the
very long quiescent interval preceding the two following outbursts.
Since the Kelvin time of the heated upper envelope is of order 1-3
months, then the 210 day cooling interval could have led to a disruption
of the time-averaged temperature equilibrium between accretional heating
and radiational cooling.  We have also uncovered the first evidence for
subsolar metal abundances in U Gem, with C markedly depleted by 0.05
with respect to the solar value, precisely what is expected to be the
processed aftermath of a C-depleting, N-enriching CNO thermonuclear
runaway.

We confirm the low white dwarf rotational velocity derived in an earlier
GHRS study (Sion et al. 1994) and find little spectroscopic evidence to
support an accretion belt on the white dwarf following outburst.  The N
V absorption lines are clearly not associated with a formation in the
white dwarf photosphere.  Our GHRS Si III data obtained near the orbital
quadratures have provided a white dwarf velocity semi-amplitude of 107 
km s$^{-1}$ and a gravitational redshift of 77 km s$^{-1}$ which
corresponds to a white dwarf mass of 1.1 M$_\odot$, if the core is made
of carbon.  A mass value this high for the U Gem degenerate is supported
by mass determinations based upon the velocity amplitudes of disk
emission lines (Stover 1981; Webbink 1990).  This high value is also not
unexpected for a CV degenerate above the period gap and suggests that,
if core mass erosion proceeds with each nova outburst, then the U Gem
degenerate must be relatively young or its mass would have been lowered.
On the other hand, if the core does not erode, then a lower limit to the
age of the system as a CV is the white dwarf cooling time (Sion 1991).
The lowest T$_{eff}$ measured for the white dwarf is T$_{eff}$ =
27,000K
(Long et al. 1996) yielding a lower limit of 5$\times10^7$ years for the
age of U Gem as a CV.

One of us (EMS) wishes to express his sincere gratitude to Dr. K.S.
Cheng and the Sir Robert Black College of the University of Hong Kong
for their kind hospitality and support during the completion of this
manuscript. This work was supported by NASA through
grant GO5412.01-94A (to Villanova University) from the Space Telescope
Science Institute, which is operated by the Association of Universities
for Research in Astronomy, Inc., under NASA contract NAS5-26555.
Partial support was also provided by NASA LTSA grant NAGW-3726 and NSF
grant AST90-16283, both to Villanova University, and by NASA LTSA grant
NAGW-3158 to the University of Washington.

\references

\noindent Cheng, F.H., Sion, E.M., Szkody, P., \& Huang, Min, 1997, AJ,
September issue

\noindent Friend, M.T., Martin, J.S., Smith, R.C., \& Jones, D.H.P.
					1990, MNRAS, 246, 637

\noindent G\"{a}nsicke \& Beuermann, K. 1996,A\&Ap, 309, L47

\noindent Livio 1993, SASSFEE Conf., eds. M. Livio, S. Shore

\noindent Long, K., Sion, E.M., Huang, M., \& Szkody, P. 1994, ApJ, 424,
					L49

\noindent Long, K., Raymond, J., Blair, W., Szkody, P., Mattei, J. 1996,
					ApJ, 469, 841

\noindent  Marsh, T.R., Horne, K., Schlegel, E.M., Honeycutt, R.K., \& 
Kaitchuck, R.H.

1990, ApJ, 364, 637

\noindent Sion, E.M., Long, K.S., Szkody, P., \& Huang, M. 1994, ApJ,
				430, L53

\noindent Sion, E.M. 1991, AJ, 102, 295

\noindent Sion, E.M. 1995, ApJ, 438, 876

\noindent Sion, E.M., Cheng, F.H. Szkody, P., Huang, M., Sparks, W.,Hubeny, I., 1996, 
				
ApJ, 480, L17

\noindent Sparks, W.M., Kutter, S., Starrfield, S., and
Truran, J. 1990, in The Physics of 

Classical Novae, ed. A. Cassatella

\noindent
Starrfield, S.G., Sparks, W.M., \& Shaviv, G. 1988,
ApJL, 326, L35

\noindent Starrfield, S.G. 1995, in Phys. Processes in Astrophys., 
ed. I. Roxborough 

(Springer: Berlin), p.99

Starrfield, S.G.,  Truran, J.,  Sparks, W.M. and Kutter, S. 1972,
ApJ, 176, L69

\noindent Stover, R. 1981, ApJ, 248, 684

\noindent Wade, R. 1981, ApJ, 246, 215

\noindent Webbink, R.1990, in Accretion-Powered Compact Binaries, ed. C.Mauche

\noindent Wood, M. 1997, unpublished model grid.

\noindent Zhang, E.-H. \& Robinson, E.L. 1987, ApJ, 321, 813

\bigskip

\newpage
\begin{center}
{\bf Figure Captions}\\
\end{center}

\noindent Fig. 1 - The AAVSO light curve data (visual magnitude versus 
time)
showing the placement of the HST observations during quiescence.
	\bigskip

\noindent Fig. 2 - The GHRS observations of U Gem in the regions of N V
(1238\AA, 1242\AA), Si III (1300 \AA), and He II (1640 \AA) during obs1
(the solid curve) and during obs2 (the dotted curve), displayed as flux
F$_{\nu}$ (mJy) versus wavelength (see text for details).
	\bigskip

\noindent Fig. 3 - The best rotating white dwarf model fit to
the three combined GHRS wavelength regions at 13 days post-outburst (top
panel).  The model fluxes are shown in bold face and span the
wavelength range 1150 \AA\ to 1650 \AA\ while the limited range of the 
three
GHRS spectra are shown with a lighter shade.  The individual GHRS G160M
regions for N V, Si III and He II are shown in the bottom panels.  Note
that the observed N V features cannot be accounted for by the white
dwarf fluxes.  Note also the photospheric C III $\lambda$1247
absorption feature just longward of N V $\lambda$1242.
	\bigskip

\noindent Fig. 4 - The same as Fig. 3, but for the observations at 61 
days post-outburst.\\

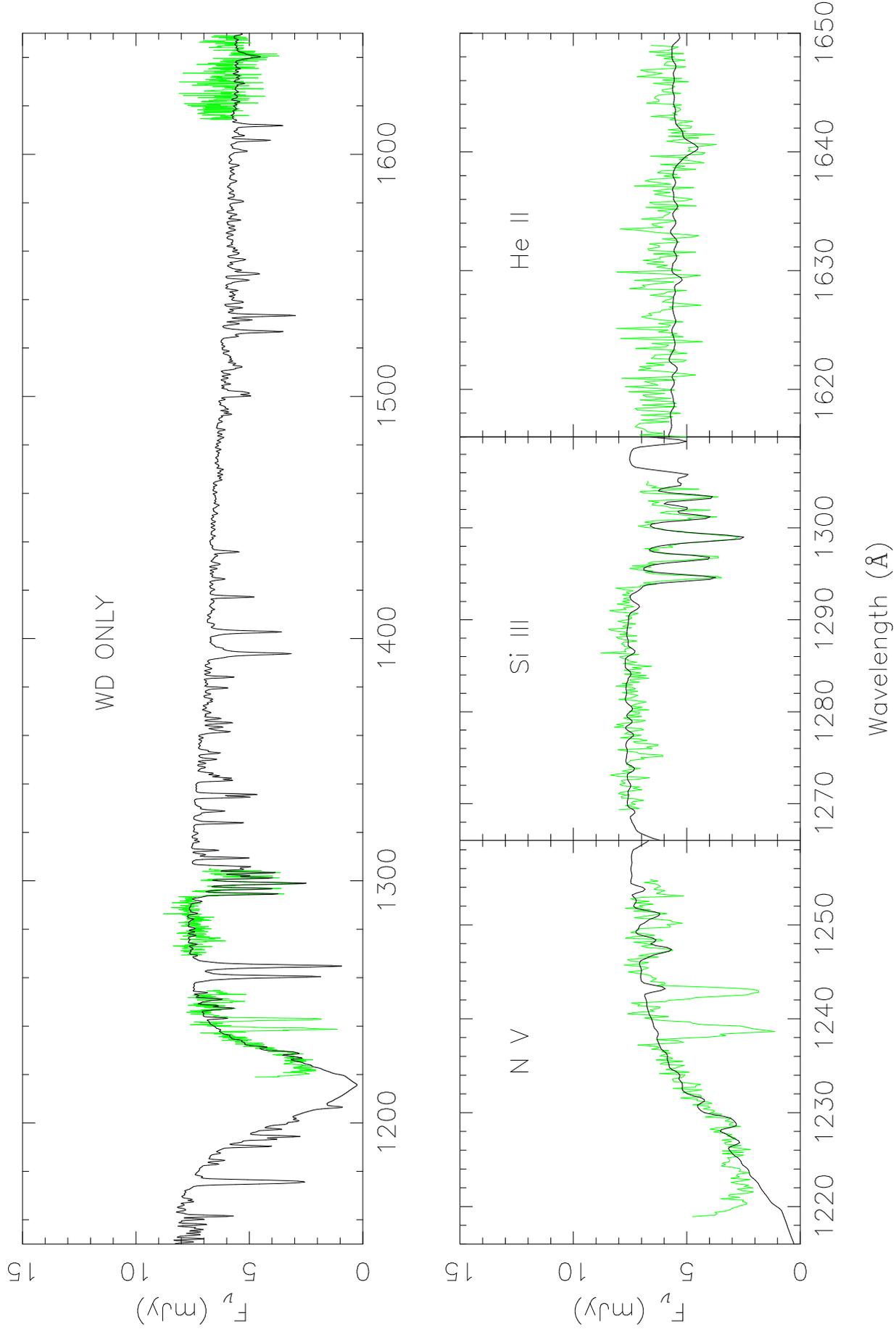
\end{document}